\documentclass[twocolumn,tighten]{aastex7}
\usepackage{graphicx}
\usepackage{amsmath}
\usepackage{scalerel}
\usepackage{apjfonts}
\usepackage{tikz}
\usetikzlibrary{svg.path}

\usepackage{comment}
\usepackage{array,multirow,color,tablefootnote}
\usepackage{xcolor}

\newcommand{\um}{$\mu$m}
\newcommand{\cotwo}{CO\textsubscript{2}}
\newcommand{\water}{H\textsubscript{2}O}
\newcommand{\methanol}{CH\textsubscript{3}OH}
\newcommand{\sal}{Salacia}
\newcommand{\man}{M{\'a}ni}

\begin{document}

\title{JWST/NIRSpec Observations of Salacia--Actaea and M{\'a}ni: Exploring Population-level Trends among Water-ice-rich Kuiper Belt Objects}


\author[orcid=0000-0001-9665-8429]{Ian~Wong}
\affiliation{Space Telescope Science Institute, 3700 San Martin Drive, Baltimore, MD, USA}
\email{iwong@stsci.edu}

\author[orcid=0000-0002-6117-0164]{Bryan~J.~Holler}
\affiliation{Space Telescope Science Institute, 3700 San Martin Drive, Baltimore, MD, USA}
\email{bholler@stsci.edu}

\author[orcid=0000-0001-8541-8550]{Silvia~Protopapa}
\affiliation{Southwest Research Institute, 1301 Walnut Street, Suite 400, Boulder, CO, USA}
\email{silvia.protopapa@swri.org}  

\author[orcid=0000-0003-2354-0766]{Aurélie~Guilbert-Lepoutre}
\affiliation{LGL-TPE, CNRS, Université Lyon 1, ENSL, Villeurbanne, France}
\email{aguilbertlepoutre@gmail.com}

\author[orcid=0000-0002-8296-6540]{William~M.~Grundy}
\affiliation{Lowell Observatory, 1400 W Mars Hill Rd., Flagstaff, AZ, USA}
\affiliation{Northern Arizona University, Flagstaff, AZ, USA}
\email{joshua.emery@nau.edu}
\email{w.grundy@lowell.edu}

\author[orcid=0000-0003-2434-5225]{John~A.~Stansberry}
\affiliation{Space Telescope Science Institute, 3700 San Martin Drive, Baltimore, MD 21218, USA}
\email{jstans@stsci.edu}

\author[orcid=0000-0001-8751-3463]{Heidi~B.~Hammel} 
\affiliation{Association of Universities for Research in Astronomy, Washington, DC, USA}
\email{hbhammel@aura-astronomy.org}

\author[orcid=0000-0001-7694-4129]{Stefanie~N.~Milam}
\affiliation{NASA Goddard Space Flight Center, Greenbelt, MD, USA}
\email{stefanie.n.milam@nasa.gov}

\author[orcid=0000-0003-3001-9362]{Rosario~Brunetto}
\affiliation{ Université Paris-Saclay, CNRS, Institut d’Astrophysique Spatiale, Orsay}
\email{rosario.brunetto@universite-paris-saclay.fr}

\author[orcid=0000-0001-9265-9475]{Joshua~P.~Emery}
\affiliation{Northern Arizona University, Flagstaff, AZ, USA}
\email{joshua.emery@nau.edu}

\author[orcid=0000-0003-2132-7769]{Estela~Fernández-Valenzuela}
\affiliation{Florida Space Institute, University of Central Florida, Orlando, FL, USA}
\email{estela@ucf.edu}

\author[orcid=0000-0002-2770-7896]{Noemí~Pinilla-Alonso}
\affiliation{Institute of Space Science and Technology of Asturias (ICTEA), University of Oviedo, Asturias, Spain}
\affiliation{Department of Physics, University of Central Florida, Orlando, FL, USA}
\email{no155980@ucf.edu}

\begin{abstract}
We present observations of the midsized Kuiper belt objects (KBOs) Salacia--Actaea and M{\'a}ni, obtained with the Near-Infrared Spectrograph on JWST. The satellite Actaea was fully blended with Salacia at the spatial resolution of the integral field unit, and we extracted the combined spectrum. The 0.7--5.1\,\um\ reflectance spectra of Salacia--Actaea and M{\'a}ni display prominent water-ice absorption bands at 1.5, 2, 3, and 4--5\,\um. The $\nu_3$ fundamental vibrational band of carbon dioxide ice at 4.25\,\um\ is present in both spectra. From a quantitative band-depth analysis of the entire current JWST spectroscopic sample of water-ice-rich KBOs, we find strong evidence for a positive covariance between relative water-ice abundance and size, which may indicate the emergent impacts of internal differentiation and cryovolcanic production of surface water ice on midsized KBOs. A detailed look at the distribution of 2 and 3 \um\ band depths suggests additional sources of variability, such as different water-ice grain sizes. In addition, we report an apparent transition in the carbon dioxide band depth at object diameters of roughly 300--500\,km, with larger objects showing systematically weaker absorptions, although selection effects within the sample do not allow us to confidently distinguish between a size-dependent phenomenon and a correlation with dynamical class. 
\end{abstract}


\section{Introduction}
\label{sec:intro}
\setcounter{footnote}{0}

Over the past three decades, observational exploration of the Kuiper belt has unveiled an astonishing diversity of surface compositions. While the reflectance spectra of the largest Kuiper belt objects (KBOs) indicate high-albedo surfaces rich in hypervolatile ices (e.g., methane, nitrogen, and carbon monoxide), early observations of smaller KBOs, at wavelengths spanning the visible and near-infrared through 2.5\,\um, revealed significantly darker regoliths and mostly featureless spectra, with only water (\water) and/or methanol (\methanol) ice being robustly detected on a handful of objects (see \citealt{barucci2008} and \citealt{brown2012review} for high-level reviews of these previous spectroscopic detections). Those results pointed toward a major transition in the surface properties of KBOs separating the largest dwarf planets (e.g., Pluto, Eris, Makemake) from the rest of the population. Volatile loss through atmospheric escape has been proposed as a first-order explanation for the absence of hypervolatiles on all but the largest KBOs. Without sufficient gravity to sustain a tenuous atmosphere and enable stable gas–solid exchange, highly volatile species such as methane would have been depleted from the surface \citep[e.g.,][]{schaller2007volatile,brown2011,protopapa2025}.

While this simple framework successfully accounts for the fundamental distinction between dwarf planets and smaller KBOs, more extensive observations of the Kuiper belt have uncovered additional complexity in its compositional landscape. Recent large-scale photometric surveys \citep[e.g.,][]{pike2017,wong2017,bernardinelli2025} and JWST spectroscopic observations \citep{depra2025,pinillaalonso2025} of KBOs with diameters less than $\sim$800\,km have identified distinct clusters of objects that differ systematically in their spectral profiles, suggesting the presence of stark compositional gradients within the primordial outer protoplanetary disk among similarly sized planetesimals \citep[e.g.,][]{brown2011,wong2016,lisse2021}. It is evident that a more robust understanding of the physical and chemical processes that have shaped the observable properties of KBOs is critical for elucidating the evolutionary trajectories of these distant icy bodies.

Midsized objects, with diameters spanning $\sim$800--1500\,km, are key to disentangling the impacts of size and formation environment on the surface composition of KBOs. The published densities of KBOs in this diameter range, which were measured from the orbital characterization of satellite-hosting systems, display a steep rise with increasing size \citep[e.g.,][]{brown2012review,mckinnon2017,grundy2019a}. Various theories have been put forth to explain this trend, including gravitational collapse of bulk porosity \citep[e.g.,][]{bierson2019}, size- and location-dependent variations in the accreted rock--ice fraction \citep[e.g.,][]{canas2024}, and large impacts \citep[e.g.,][]{barr2016}. These objects are also an ideal testing ground for probing interior processes such as differentiation, radiogenic heating, and subsurface water-ice melt, which are expected to have a substantial impact on the internal evolution of KBOs at these sizes (see, for example, reviews by \citealt{mckinnon2008} and \citealt{guilbertlepoutre2020}).

\citet{emery2024} present 0.6--5\,\um\ JWST observations of three midsized KBOs---Gonggong, Quaoar, and Sedna---which straddle the predicted threshold for methane (CH$_{4}$) ice retention from atmospheric volatile loss models \citep[e.g.,][]{schaller2007volatile,steckloff2021,protopapa2025}. The measured spectra are markedly different from both the larger dwarf planets (Pluto, Eris, Makemake) and the smaller KBOs. The most notable feature is the presence of ethane (C$_{2}$H$_{6}$) ice on all three objects, with Sedna showing additional light hydrocarbon signatures from ethylene and acetylene. Meanwhile, the spectrum of Quaoar yields a clear detection of CH$_{4}$ ice, confirming previous findings from ground-based data \citep{schaller2007quaoar,dalleore2009}. 

Given that C$_{2}$H$_{6}$ and other light hydrocarbons are the immediate byproducts of CH$_{4}$ irradiation \citep[e.g.,][]{gerakines2001,bennett2006}, and that they in turn readily photolyze into more complex refractory hydrocarbons \citep[e.g.,][]{moroz2004,bennett2006,quirico2023,zhang2023}, their persistence on the surfaces of these midsized KBOs likely requires steady replenishment of CH$_{4}$, perhaps from a geochemically sourced reservoir in the interior of these bodies \citep{glein2024,grundy2024,protopapa2025}, along with a sufficiently long residence time on the surface to allow for radiation processing \citep{emery2024}. These findings suggest that midsized KBOs are not quiescent, primitive bodies, but rather dynamic worlds that manifest both past and possibly ongoing endogenic and exogenic alteration---including internal differentiation, surface--interior exchange, insolation-driven volatile cycling, and space weathering.

In this paper, we present observations of \sal--Actaea \citep{roe2005,noll2006} and \man\ \citep{trujillo2002}\footnote{Alternate designations: \sal\ = 120347 (2004 SB60), \man\ = 307261 (2002~MS4).} obtained with the Near-Infrared Spectrograph (NIRSpec) on JWST. With radiometric diameters of $901\pm45$ and $934\pm47$\,km \citep{vilenius2012,fornasier2013,muller2020}, respectively, these two objects bridge the gap in size between the three midsized objects presented in \citet{emery2024} and the sample of smaller KBOs characterized in \citet{pinillaalonso2025}. Hubble imaging of \sal\ revealed the binary companion Actaea \citep{noll2008}, which has an orbital period of $\sim$5.5\,days \citep{grundy2019}. Using the measured primary--secondary brightness difference of $2.37\pm0.06$\,mag \citep{stansberry2012} and assuming the same albedo, the diameter of Actaea is estimated to be $290\pm21$\,km \citep{brown2017}. The JWST observations do not resolve the two components of the binary system, and we analyze the combined spectrum. Hereafter, we refer to the Salacia--Actaea system as \sal, for the sake of brevity. 

Despite being among the 10 largest known KBOs, information about the surface properties of \sal\ and \man\ is relatively scant in the published literature. Ground-based spectroscopy of \sal\ showed a moderately red visible color \citep{pinillaalonso2008} and a featureless near-infrared spectrum from 1.4 to 2.4\,\um\ \citep{schaller2008}. Meanwhile, no published spectroscopy of \man\ exists; broadband visible photometry of \man\ indicated a relatively neutral spectral slope, similar to that of Haumea \citep{tegler2016}. A stellar occultation of \man\ revealed a plausibly impact-related topographic depression and an area-equivalent diameter of $796\pm24$\,km---significantly smaller than the radiometric measurement, suggesting a possible thermal contribution from an undiscovered dark satellite \citep{rommel2023}. The JWST data described in this work provide the first detailed view of \sal\ and \man\ at infrared wavelengths.

\section{Observations and Data Analysis}
\label{sec:obs}

Observations of Salacia and \man\ were carried out with the integral field unit (IFU) of NIRSpec \citep{jakobsen2022,boker2023} on UT 2022 November 12 and 2022 August 31, respectively, as part of JWST Cycle 1 Guaranteed Time Observations Program \#1191 (PI: J. Stansberry). Each observation consisted of a pair of dithered exposures using the low-spectral-resolution double-pass prism, which provides continuous wavelength coverage from 0.6 to 5.3\,\um\ at a spectral resolution of $\Delta\lambda/\lambda = 30-300$. Each exposure had a total integration time of 759\,s and utilized the NRSIRS2RAPID readout method, which reduces the effect of detector read noise \citep{moseley2010,rauscher2012}. The heliocentric distance ($r_h$), distance from JWST ($d$), and phase angle ($\alpha$) of Salacia and \man\ at the time of the observations were $r_h = 45.1$\,au, $d = 44.5$\,au, $\alpha = 0\overset{\circ}{.}94$ and $r_h = 46.3$\,au, $d = 45.7$\,au, $\alpha = 1\overset{\circ}{.}02$, respectively.

\begin{figure}
    \includegraphics[width=\columnwidth]{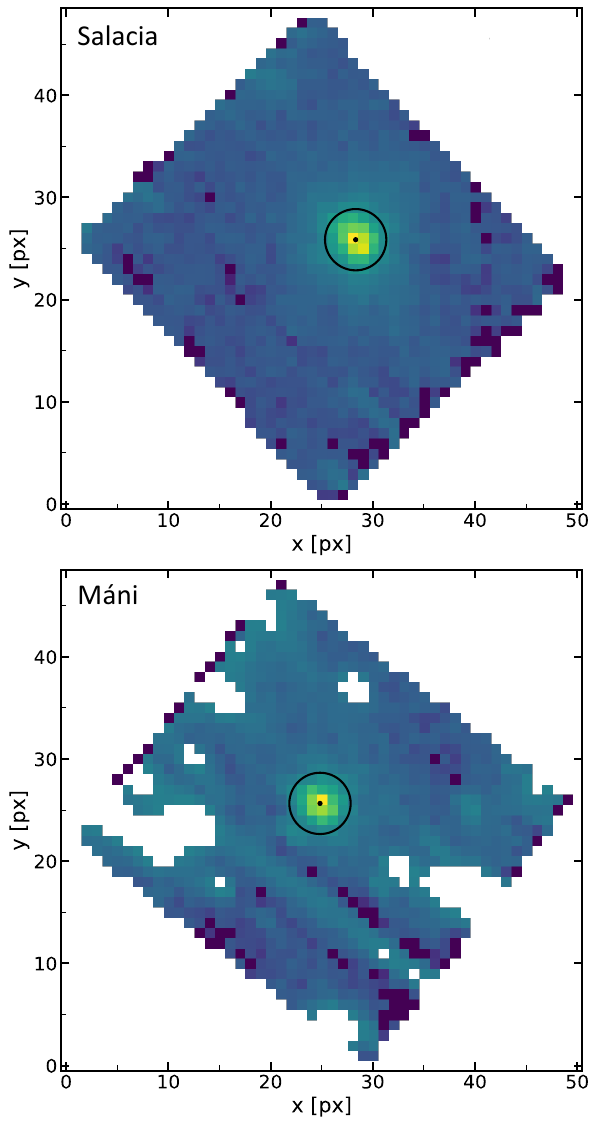}
    \caption{Wavelength-averaged IFU images from the first dithered exposures of \sal\ and \man. A logarithmic stretch was applied to the flux scaling to highlight the PSF shapes and background variations. The black dots and circles denote the best-fit centroid positions and the spectral extraction apertures, respectively. In the case of \man, the masked pixels across the field of view correspond to relatively bright background sources that were flagged prior to spectral extraction (see text for details).}
    \label{fig:images}
    \vspace{-0.1cm}
\end{figure}

Starting with the raw uncalibrated detector images downloaded from the Mikulski Archive for Space Telescopes, we used the \texttt{jwstspec} pipeline \citep{jwstspec} to generate dark-corrected, flat-fielded, readnoise-corrected, spatially-rectified, and flux-calibrated data cubes for each dithered exposure and carried out optimal spectral extraction; a full description of \texttt{jwstspec} and its implementation for spectroscopic observations of solar system small bodies is provided in \citet{wong2025} and references therein. The core of the data processing workflow is the official JWST calibration pipeline, and the results presented in this paper were derived using Version 1.17.1 \citep{jwst}; all necessary reference files were drawn from context \texttt{jwst\_1321.pmap} of the JWST Calibration Reference Data System. 

Figure~\ref{fig:images} shows the wavelength-averaged IFU images from the first dithered exposures of Salacia and \man. A visual inspection of the point-spread function (PSF) of Salacia reveals a slight deviation from a radially symmetric shape. At the time of the observation, Actaea was $0\overset{''}{.}1$ to the northeast of \sal, according to the JPL Horizons ephemeris, which corresponds to one NIRSpec IFU spatial pixel. It follows that the PSFs of \sal\ and Actaea are fully blended in the IFU cubes, and we proceeded to extract both sources together. Assuming the same albedo across the binary pair, the diameter of Actaea is approximately one-third that of Salacia \citep{stansberry2012,brown2017}, corresponding to $\sim$10\% of the total system flux---a negligible contribution to the measured spectrum. In the case of \man, the target was located in a dense field of background stars. To ensure a reliable measurement of the background flux level, we masked the target and applied an iterative $\sigma$-clipping filter to the median image of each data cube to exclude pixels that exceeded the median by more than $3\sigma$. 

\begin{figure*}
    \includegraphics[width=\textwidth]{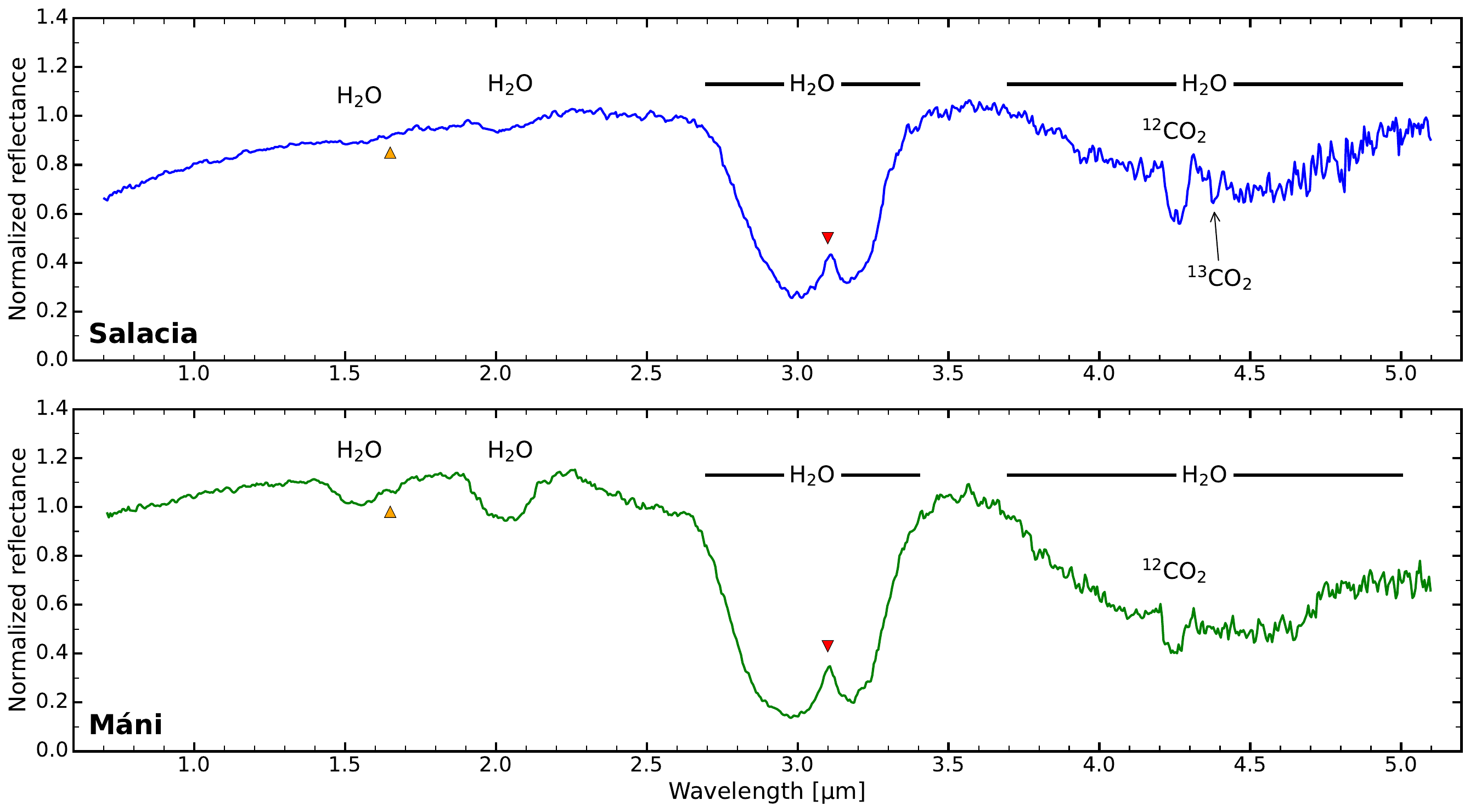}
    \caption{The reflectance spectra of \sal\ (top) and \man\ (bottom), normalized to unity at 2.5\,\um. The detected absorption bands of \water\ and \cotwo\ ice are labeled. The red triangles correspond to the location of the 1.65\,\um\ \water-ice spectral feature that likely indicates the crystalline phase, which is clearly discernible in the spectrum of \man. The inverted orange triangles mark the 3.1\,\um\ Fresnel reflection peak of \water\ ice.}
    \label{fig:reflectance}
\end{figure*}

We calculated the target's centroid in each data cube from a Gaussian fit to the source PSF and extracted the spectrum using the empirical PSF fitting method within \texttt{jwstspec}. Template PSF models were generated for each slice using a 21 pixel wide moving median window along the wavelength axis; these templates were then normalized and fit to the slices with a multiplicative scaling factor to determine the fluxes. Uncertainties were estimated by weighting the pipeline-generated error array of each slice with the corresponding normalized template PSF and computing the quadrature sum of values within the extraction aperture. For both targets, the radius of the spectral extraction aperture was set to 3 pixels, and the background was measured across all unmasked pixels outside of a $21\times21$\,pixel box centered on the target's centroid. The spectra from the pairs of dithered exposures were passed through a 21 point wide $5\sigma$ moving median filter to remove outliers and then averaged using a simple mean to produce the combined irradiance spectrum. The associated uncertainties were obtained through standard error propagation.

To derive the reflectance spectra and account for common-mode instrumental systematics, the irradiance spectrum of each target was divided by the analogously derived spectrum of the solar analog star SNAP-2 (Program \#1128; PI: N. Luetzgendorf). To correct for the spectral difference between SNAP-2 and the Sun, we multiplied the reflectance spectrum by the wavelength-dependent ratio between the solar standard spectrum from the Planetary Spectrum Generator \citep{psg} and the model spectrum of SNAP-2 from CALSPEC \citep{bohlin2014}. Figure~\ref{fig:reflectance} shows the final normalized reflectance spectra of \sal\ and \man. Both spectra were trimmed to 0.7--5.1\,\um\ to remove the wavelength regions of low instrument throughput and poor data quality.

\section{Discussion}
\label{sec:disc}

\subsection{Spectral Characterization}\label{subsec:char}

The spectra of \sal\ and \man\ have similar shapes. Both show a deep absorption feature at 3\,\um, a broad reflectance minimum spanning 4--5\,\um, and weaker bands at 1.5 and 2\,\um---all of which are the characteristic spectral signatures of \water\ ice. In addition, the two spectra display a sharp Fresnel reflectance peak at 3.1\,\um\ (marked with the red triangles in Figure~\ref{fig:reflectance}). At 1.65\,\um, there is a distinct absorption band that, together with the 3.1\,\um\ Fresnel peak, is clearly discernible on \man\ (orange triangle); for \sal, it is difficult to ascertain whether a similar 1.65\,\um\ \water-ice feature is present, given the spectrum's weak 1.5\,\um\ absorption band. Overall, the \water-ice features on \man\ are significantly stronger than on \sal\ across the observed wavelength range, suggesting a higher relative abundance of \water\ on its surface. There are also some differences in the shape of the \water-ice features between the two objects: for example, the profile of the broad 4--5\,\um\ absorption band on \man\ has a distinctively sharp rise at around 4.7\,\um, which is not seen on \sal. An examination of \water-ice optical constants indicates systematic variations in the 4--5\,\um\ band shape between the crystalline and amorphous phases, with the former showing a somewhat steeper and shorter-wavelength decrease in absorbance across the 4.5--5.0\,\um\ region \citep{mastrapa2009}---broadly consistent with the observed spectral shape on \man. It should be noted, however, that variations in \water-ice path length as well as surface layering can also contribute to differences in band shapes, relative band depths, and the presence/absence of the Fresnel peak; proper radiative transfer modeling, which is outside the scope of this paper, is required to fully explore the effects of abundance, grain size, and crystallinity \citep[e.g.,][]{hansen2004,protopapa2014,protopapa2021}. 

At 4.25\,\um, both \sal\ and \man\ show a pronounced absorption band attributed to the $\nu_3$ asymmetric stretch mode of solid carbon dioxide (\cotwo; e.g., \citealt{gerakines2020}). \sal\ has an additional narrow feature at 4.37\,\um\ that corresponds to the $^{13}$\cotwo\ isotopologue \citep[e.g.,][]{he2018}. This feature is not apparent on \man, although we note that the typical signal-to-noise ratio of the spectra in this region is relatively low compared to shorter wavelengths, which makes the detection of weak, narrow features such as the $^{13}$\cotwo\ band very challenging. Indeed, the overall scatter in measured reflectance for \sal\ across 4.3--4.7\,\um\ is comparable to the observed depth of the $^{13}$\cotwo\ feature. Future JWST observations with significantly longer exposure times are needed to precisely measure the $^{13}$C/$^{12}$C ratio in \cotwo\ ice on KBOs and probe for trends across the population.

Spectroscopic observations of dozens of smaller KBOs obtained with NIRSpec during the first year of JWST science operation have revealed three compositional classes \citep{depra2025,licandro2025,pinillaalonso2025}: (1) a subpopulation with moderately red visible colors that displays strong \water-ice absorption bands and weaker \cotwo-ice features; (2) a class with systematically redder colors and deep \cotwo-ice features that also exhibits prominent absorptions from solid carbon monoxide and enhanced signatures of irradiated aliphatic organics; (3) a separate group objects with even redder average visible colors that display a negative-sloped near-infrared continuum and much stronger organic features, along with the diagnostic spectral bands of \methanol\ ice. Members of the Haumea collisional family are considered a separate group of exceptionally water-ice-rich KBOs and are not included as part of the three compositional classes delineated here. Recently, \citet{holler2025rnaas} proposed a new taxonomic nomenclature for these three compositional classes---spectrally prominent water (\water-type), spectrally prominent carbon oxides (\cotwo-type), and spectrally prominent organics (organics-type)---which we adopt in this paper. These classes have previously been referred to, respectively, as ``bowl''-, ``double-dip''-, and ``cliff''-type in the literature \citep[e.g.,][]{pinillaalonso2025}. 

\begin{figure}[t!]
    \includegraphics[width=\columnwidth]{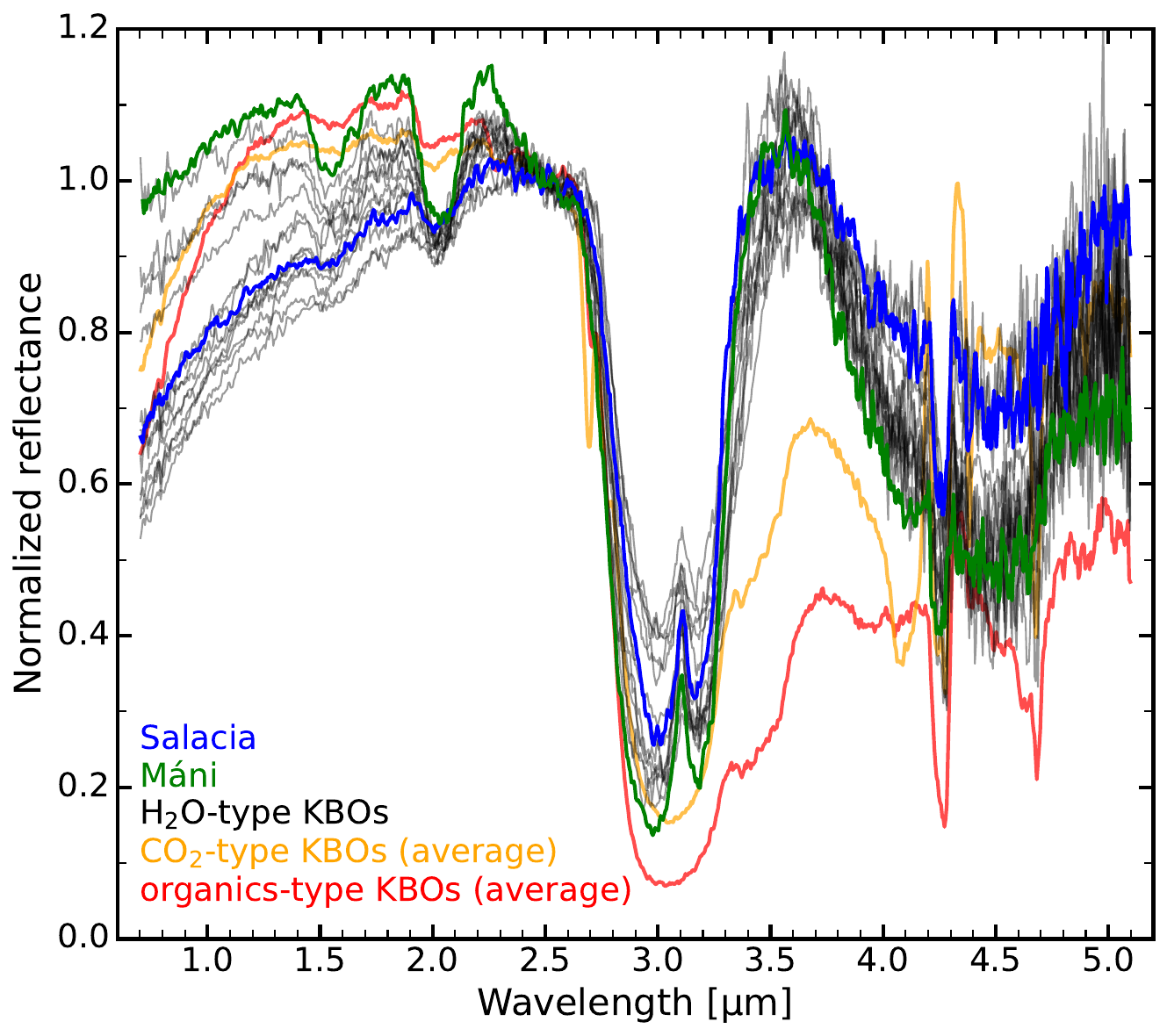}
    \caption{The reflectance spectra of \sal\ (blue) and \man\ (green) alongside the sample of \water-type KBOs (black) from \citet{pinillaalonso2025}, illustrating the broad similarities in overall continuum shape and absorption bands. The average \cotwo-type KBO (orange) and organics-type KBO (red) spectra are also included for comparison. All spectra have been normalized to unity at 2.5\,\um.}
    \label{fig:discocompare}
\end{figure}

Visual comparisons of \sal\ and \man\ with the three compositional classes show that their spectra align most closely with the \water-type KBOs. In Figure~\ref{fig:discocompare}, the spectra of \sal\ and \man\ are plotted against the set of individual spectra of \water-type KBOs from \citet{pinillaalonso2025}, along with the average spectra of the \cotwo-type and organics-type KBOs. All of the \water-type KBO spectra exhibit the same prominent \water\ and \cotwo\ absorption bands and overall spectral profile as \sal\ and \man. Meanwhile, the \cotwo- and organics-type spectra have drastically different continuum shapes and distinct collections of absorption features; most notably, both \cotwo- and organics-type KBOs show strong aliphatic organics signatures at 3.3--3.6\,\um\ \citep[e.g.,][]{brunetto2006}, which are not seen on \sal, \man, or any of the other \water-type KBOs.

The stark spectral differences between the aforementioned three  compositional classes suggest distinct formation locations in the outer solar system. Several previous works have hypothesized the existence of steep compositional gradients within the primordial trans-Neptunian planetesimal disk, driven by differential volatile retention as a function of heliocentric distance \citep[e.g.,][]{brown2011,wong2016,lisse2021}. It follows that the \water-type KBOs, which show the fewest volatile ice signatures, would have formed closest to the inner edge of this disk \citep{pinillaalonso2025}. Within this explanatory framework, \sal\ and \man\ can be classified as two of the largest members of the broader \water-type class that all shared a common formation environment and were subsequently scattered throughout the Kuiper belt during the dynamical instability \citep[e.g.,][]{levison2008,nesvorny2018}.

Notably, the nondetection of the diagnostic spectral signatures of simple hydrocarbons (e.g., CH$_{4}$ and C$_{2}$H$_{6}$) and the absence of the broad 3.3--3.6\,\um\ absorption band due to aliphatic organics on \sal\ and \man\ mark a critical distinction with Gonggong, Quaoar, and Sedna \citep{emery2024}, which are only slightly larger than \sal\ and \man, with radiometric diameters of $1230\pm50$\,km \citep{kiss2019}, $1074\pm38$\,km \citep{fornasier2013}, and $995\pm80$\,km \citep{pal2012}, respectively. One possible explanation for this discrepancy is that the smaller size and consequently weaker surface gravity of \sal\ and \man\ do not allow for a sufficiently long retention time of CH$_4$ ice to enable the production of longer-chain hydrocarbons through radiolysis or photolysis. To explore this line of reasoning more quantitatively, we consider the escape velocities of midsized KBOs. For \sal, Quaoar, and Gonggong, which have masses constrained by observations of their satellites \citep[e.g.,][]{grundy2019,kiss2019,collyer2025,proudfoot2025}, the escape velocities are 385, 542, and 616\,m\,s$^{-1}$, respectively. The speed of a CH$_4$ molecule desorbed from a surface with a temperature of 30--50\,K is governed by the Armand distribution \citep[e.g.,][]{armand1977}, which peaks at lower velocities in the 200--300\,m\,s$^{-1}$ range. Only the most rapidly moving molecules in the tail of the Armand distribution have sufficient momentum to escape. The difference between escape velocities of 385 and 542\,m\,s$^{-1}$ translates to a roughly order of magnitude larger fraction of desorbed CH$_4$ molecules that exceed the escape velocity on \sal\ than on Quaoar. 

It follows that a CH$_4$ molecule on Quaoar's surface undergoes an order of magnitude more ballistic hops before escaping. This greatly increases the probability that a CH$_4$ molecule encounters a cold trap (e.g., the winter hemisphere of a high-obliquity object), where it is stable to sublimation loss. \citet{grundy2016} and \citet{raut2022} demonstrated that resonantly scattered Ly$\alpha$ photons can strike the nonilluminated hemispheres of KBOs in sufficient quantities to effectuate irradiation processing of cold-trapped CH$_4$ into heavier, less volatile hydrocarbons (e.g., C$_{2}$H$_{6}$). The situation is more uncertain for \man\ and Sedna, which are very close in size but have no dynamical mass measurements. If we take their radiometric diameters and assume the same bulk density as Salacia---1.50\,g\,cm$^{-3}$ \citep{grundy2019}---we obtain comparable escape velocities of 430 and 460\,m\,s$^{-1}$, respectively. However, Sedna's distant aphelion entails lower surface temperatures than the other midsized KBOs, which consequently augment its ability to retain CH$_{4}$ and produce the light hydrocarbon signatures seen in its spectrum.

It is also possible that the absence of C$_{2}$H$_{6}$ or other light hydrocarbons is due to a much lower inherent abundance of CH$_4$ on \sal\ and \man\ than on Gonggong, Quaoar, and Sedna. In light of the proposed geochemical origin of endogenically sourced CH$_4$ on the dwarf planets Eris and Makemake \citep{glein2024,grundy2024}, the absence of CH$_4$ and its associated irradiation products may indicate that \sal\ and \man\ do not have the requisite physical conditions in their interiors to support the thermogenic production of CH$_4$ and its transport to the surface. Alternatively, based on the previously discussed explanatory framework for the three KBO compositional classes, the broad similarity between \sal, \man, and the \water-type KBOs suggests that they formed in the innermost region of the primordial trans-Neptunian disk, where the higher temperatures led to rapid depletion of CH$_4$ ice on the surfaces of planetesimals, thereby curtailing the formation of hydrocarbons and more complex refractory organics that could be incorporated into the resultant bodies.

\subsection{Trends in Absorption Band Depths} \label{subsec:trends}

The ensemble of \water-type KBO spectra in Figure~\ref{fig:discocompare} spans a broad range of relative band depths and continuum slopes. In fact, regardless of the choice of normalization wavelength, these spectra exhibit a significantly larger dispersion in spectral shape than the PO- and \cotwo-type compositional classes, indicating a particularly high level of compositional diversity within the \water-type class. This finding motivated us to examine the variations in spectral properties among \water-type KBOs in a more quantitative manner. 

We carried out a systematic band-depth analysis of the strongest \water\ and \cotwo\ features at 2, 3, and $\sim$4.25\,\um. The spectroscopic sample was expanded to include other \water-type targets that were observed with NIRSpec: (a) six Neptune Trojans that were published in \citet{markwardt2025}, (b) four Centaurs from \citet{licandro2025}, (c) three extreme trans-Neptunian objects (ETNOs) with distant perihelia observed as part of Cycle 3 Program \#4665 (Holler et al. 2025, in preparation), (d) the blue binary cold classical KBO 2016~BP81 \citep{wong2025}, (e) the midsized objects Orcus and Achlys, from Cycle 1 Guaranteed Time Observations Program \#1231, and (f) Charon, which was presented in \citet{protopapa2024}. The full sample includes 33 objects. All spectra were extracted using an analogous procedure to the one applied to \sal\ and \man\ (Section~\ref{sec:obs}). When computing the band depths and associated uncertainties, we followed the methodology described in \citet{wong2025}; the same continuum and band center regions were used, with the exception of the \cotwo\ fundamental band near 4.25\,\um. For some of the largest objects---namely, Orcus, \man, and Achlys---the location of minimum reflectance is shifted toward shorter wavelengths, and we correspondingly altered the designated band center region from 4.26--4.28\,\um\ to 4.24--4.26\,\um\ when calculating the \cotwo\ band depth for those three objects. The implications of this spectral behavior are discussed in the following subsection.

\begin{figure}[t!]
    \includegraphics[width=\columnwidth]{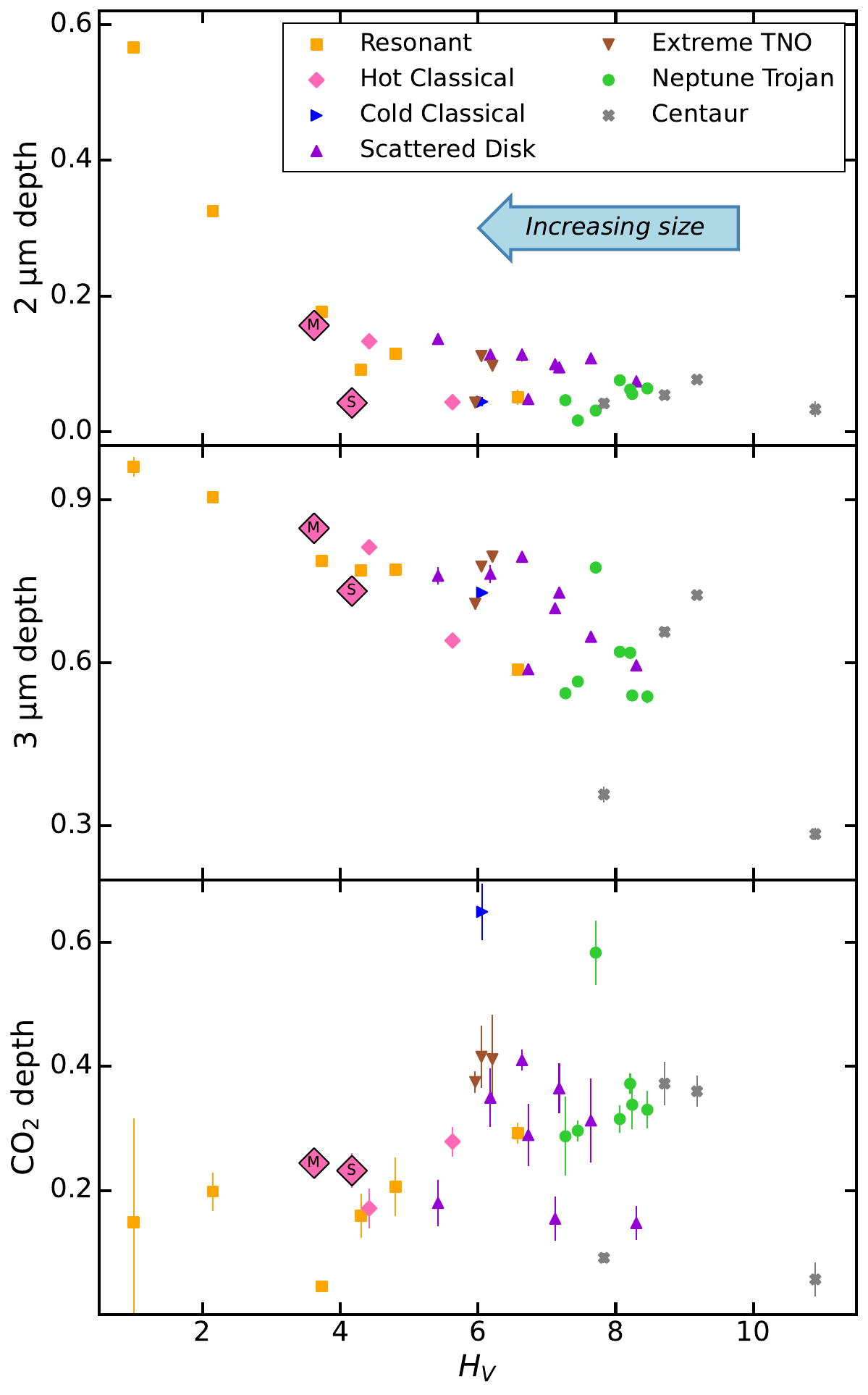}
    \caption{Plot of the measured band depths of the 2\,\um\ \water\ (top), 3\,\um\ \water\ (middle), and $\sim$4.25\,\um\ fundamental \cotwo\ absorption features as a function of $V$-band absolute magnitude $H_{V}$ for \water-type objects. The points are color coded by dynamical class. The band depths for \sal\ and \man\ are highlighted by the larger markers and labeled ``S'' and ``M,'' respectively. The three brightest/largest \water-type members of the resonant KBO population (orange squares) are, in order, Charon, Orcus, and Achlys. The band-depth values, absolute magnitudes, and dynamical classifications shown in this figure are available as a machine-readable data behind the figure file.}
    \label{fig:stack}
\end{figure}

In Figure~\ref{fig:stack}, the measured 2, 3\,\um, and CO$_2$ fundamental band depths for the \water-type KBOs are plotted against $V$-band absolute magnitude ($H_{V}$), which we use here as a proxy for size; the $H_{V}$ values were sourced from JPL Horizons. The objects are color coded to denote their respective dynamical classifications, as tabulated in \citet{volk2024}. \sal\ and \man\ are labeled with ``S'' and ``M,'' respectively. There are clear selection effects in the dataset with respect to object brightness---the Neptune Trojans and Centaurs sample systematically higher $H_{V}$ values (i.e., smaller sizes) due to their smaller heliocentric distances, while the largest and brightest objects are exclusively from the resonant and hot classical KBO populations. However, when assessing the band-depth distribution of objects with similar $H_{V}$ values, no significant correlations with dynamical class are apparent.

\subsubsection{Water Ice}

The depth of the 2\,\um\ \water-ice absorption feature decreases with increasing $H_{V}$, indicating that smaller/fainter objects tend to show weaker \water-ice spectral signatures. The relative positions of \sal\ and \man\ in the top panel of Figure~\ref{fig:stack} are readily validated by a visual comparison of their reflectance spectra (Figure~\ref{fig:reflectance}). To assess the statistical significance of the depth~versus~$H_{V}$ trend across the full sample, we computed the Pearson correlation coefficient and obtained $r=-0.72$ with a $p$-value of $1.4\times10^{-6}$, indicating that the correlation is statistically significant at the $4.7\sigma$ level. However, the strength of the overall correlation is primarily driven by the largest objects with $H_V \lesssim 5$, where the 2\,\um\ depth decreases steeply with increasing $H_{V}$ ($r=-0.95$, $3.6\sigma$). Among smaller objects, the covariance is weaker and less statistically significant ($r=-0.35$, $1.7\sigma$). This behavior was previously reported based on analyses of ground-based spectra \citep[e.g.,][]{barucci2011,brown2012h2o}, although the precision of those band-depth measurements for objects with $H_{V} \gtrsim 4$ was severely limited by the quality of the data. Moreover, almost all of the objects with $H_{V} \gtrsim 6$ in the extant spectroscopic sample at that time were Centaurs, which may have experienced a divergent surface evolution from the KBO population due to their smaller heliocentric distances. As such, the relationship between band depth and size among KBOs with $H_{V} \gtrsim 4$ was uncertain.

The exquisite precision of the JWST spectra has provided a detailed view of the 2\,\um\ band-depth distribution through $H_{V} = 8$. Notably, there is strong evidence for a bifurcation in the measured 2\,\um\ band depths, particularly within the range $4 < H_{V} < 8$. Two subgroups are identified: one with near-zero band depths (including \sal) that shows no clear correlation with $H_{V}$ and another with systematically deeper absorption features that exhibits a modest decrease in depth with increasing $H_{V}$. Applying Hartigan's dip test \citep{hartigan1985} yields a $3.3\sigma$ detection of bimodality. 

The broad near-infrared wavelength coverage provided by JWST has uncovered, for the first time, a robust size-dependent trend in the 3-\um\ \water\ absorption band. The 3\,\um\ band depth decreases monotonically with increasing $H_{V}$ across the sample of \water-type KBOs ($r=-0.82$, $5.4\sigma$). Moreover, we find statistically significant correlations within individual dynamical classes that span more than 2\,mag in $H_{V}$, specifically resonant objects, hot classicals, and scattered disk objects. This suggests that the relationship between 3\,\um\ band depth and size is not confined to particular regions of the Kuiper belt but is instead a general feature of the population at large. 

The distinctive steepening of the 2\,\um\ band-depth trend at the smallest $H_{V}$ values is not mirrored in the 3\,\um\ band depths. This can be attributed to the onset of saturation, with the two largest KBOs in the sample---Charon and Orcus---having 3\,\um\ band depths in excess of 90\%. Meanwhile, two Centaurs---Okyrhoe and 2010~KR59---have exceptionally shallow 3-\um\ absorptions. This behavior, which is discussed in detail in \citet{licandro2025}, may stem from the effects of secondary evolutionary processes that have altered the surfaces of these Centaurs following their migration into the warmer giant planet region. Thermal processes, such as devolatilization and the formation of ice-poor lag deposits resulting from cometary activity, provide plausible explanations for the observed attenuation of the 3\,\um\ \water-ice absorption feature. Due to the possibly evolved nature of Centaurs, we do not consider them in the following discussion.

Notably, unlike in the case of the 2\,\um\ band depths, no apparent bifurcation is apparent in the 3\,\um\ band depths across the range $4 < H_{V} < 8$. This contrast may arise from variations in the extinction coefficient of \water\ ice, which is much smaller at 2\,\um\ than at 3\,\um.  Consequently, the penetration depth of reflected sunlight into the surface is more than an order of magnitude greater at 2\,\um\ \citep[e.g.,][]{protopapa2024}. The observed bimodality in the 2\,\um\ band depth among the \water-type objects could therefore be indicative of a marked difference in \water-ice grain sizes below the outermost surface layer.

If we assume that the 3\,\um\ band depth serves as a proxy for \water-ice abundance, the overall trend in the 3\,\um\ band depths is consistent with a steady increase in relative surface \water-ice abundance with increasing size. There is, however, considerable scatter in the band-depth distribution, particularly at $H_{V} > 6$. The significant spread in measured depths among similarly sized objects may be attributed to variations in \water-ice grain size across the \water-type population. Taken together, the behavior of the 2 and 3\,\um\ \water-ice absorption features may indicate some level of vertical stratification on the surfaces of these objects. Detailed spectral modeling of the full spectroscopic ensemble is necessary to retrieve reliable information about the exact physical configuration of \water\ ice on these bodies and disentangle the effects of abundance, grain size, and layering on the observed spectra.

Both radioactive decay and the onset of internal differentiation are exothermic processes whose impacts strongly scale with increasing size (see review by \citealt{guilbertlepoutre2020}). Theoretical modeling has demonstrated that differentiation of ice-rich planetesimals can begin to occur at diameters above $\sim$500\,km \citep[e.g.,][]{hussmann2006,mckinnon2008,malamud2015}, which corresponds to $H_{V}\sim 4.7$, assuming a geometric albedo of 0.1. As seen in Figure~\ref{fig:stack}, this brightness level roughly matches the location below which the 2 and 3\,\um\ \water-ice band depths rise most dramatically. Radiogenic heating due to the decay of long-lived isotopes may also be enhanced in large- and midsized KBOs above $\sim$500--1000\,km in diameter, which have systematically higher bulk densities and plausibly higher rock--ice fractions \citep[e.g.,][]{brown2012review,mckinnon2017}. The internal heating from these processes likely caused widespread melting and mobilization of \water\ in the interiors of the largest KBOs and may have sustained subsurface liquid water oceans for a significant fraction of the age of the solar system \citep[e.g.,][]{hussmann2006,mckinnon2008,desch2009}. Cryovolcanic activity could have facilitated the advection of liquid water onto the exterior \citep[e.g.,][]{cook2007,desch2009,delsanti2010,menten2022}, thereby enriching the surface with \water\ ice and producing the observed positive correlation between relative \water-ice abundance and size. Large-scale cryovolcanism has been proposed to explain the compositional and geological diversity on the surfaces of Pluto and Charon revealed by \textit{New Horizons} (see review by \citealt{moore2021} and references therein).

If the observed covariances between \water-ice band depths and $H_V$ are indicative of an enrichment of \water\ ice on the surfaces of the largest objects, we expect to find corresponding size-modulated variability in both surface color and albedo. The reflectance spectrum of \water\ ice has a distinctively negative-sloped continuum across the 0.5--2.5\,\um\ range \citep[e.g.,][]{grundy1998,warren2008,mastrapa2013}; an enhancement in relative surface \water-ice fraction should therefore manifest as a bluing of the spectral slope. Likewise, a higher surface ice fraction should entail an elevated visible geometric albedo. As illustrated in Figure~\ref{fig:discocompare}, the \water-type KBOs span a wide range of near-infrared continuum slopes. We renormalized all of the reflectance spectra in our sample to unity at 1.0\,\um\ and fit a linear slope across 0.7--1.0\,\um. While most of the measured slope values do not exhibit a correlation with $H_V$ or 2\,\um\ \water-ice band depth, the four largest objects (Charon, Orcus, Achlys, and \man) have systematically shallower (e.g., more neutral) continua than the rest of the sample. Similarly, we assessed the distribution of radiometrically measured albedo values across the \water-type sample (taken from the compilation in \citealt{pinillaalonso2025} and references therein). While typical albedos (including those of \sal\ and \man) lie between 0.04 and 0.13, the two largest objects are significantly more reflective. Orcus has a measured albedo of $0.231\pm0.018$ \citep{fornasier2013}; \textit{New Horizons} mapping of Charon's surface yielded albedos in the range 0.20--0.73 \citep{buratti2017}. These findings offer strong support to the idea of systematic surface \water-ice enrichment on the largest \water-type KBOs.

\subsubsection{Carbon Dioxide Ice}

Turning our attention now to \cotwo\ (bottom panel of Figure~\ref{fig:stack}), we find two objects---the cold classical binary 2016~BP81 and the Neptune Trojan 2011~SO277---with significantly larger $\sim$4.25\,\um\ band depths than the rest of the sample. The distinct spectral behavior of these two outliers was previously reported by \citet{wong2025}, who posited that the unique characteristics may reflect in situ formation of blue binaries within the cold classical region during the earliest stages of the protoplanetary disk, under markedly different thermophysical conditions than objects that accreted later \citep{nesvorny2022}. The subsequent orbital restructuring of the Kuiper belt would then have scattered a fraction of these objects into different dynamical populations, thereby explaining the distinctive spectral shape of 2011~SO277 relative to the other Neptune Trojans in the sample.

Excluding the two objects with exceptionally deep \cotwo-ice features (as well as the Centaurs), we find strong evidence for a transition in the average \cotwo-ice band depth at around $H_{V} \sim 5-6$, with the larger KBOs displaying systematically weaker \cotwo\ absorptions than the smaller objects. When applying the two-sample Kolmogorov--Smirnov test to the \cotwo\ band-depth distributions on either side of $H_{V}=5.5$, we found that they differ at the $4.2\sigma$ level. Meanwhile, there is no evidence for correlations with $H_{V}$ on either side of the threshold ($|r|<0.15$). Considering the selection biases among the various dynamical classes within the spectroscopic sample, it is not possible to firmly distinguish between an overall trend with size (as in the case of the \water-ice absorption features) and a dependency on dynamical class. It is worth noting that, while the resonant objects, hot classicals, Neptune Trojans, and ETNOs appear to cluster within the space of measured \cotwo\ band depths, the scattered disk objects span a much broader range of values. Additional observations of \water-type KBOs that expand the spectroscopic sample within the various dynamical classes to a wider range of sizes are needed to definitively resolve this ambiguity in the nature of the observed \cotwo\ band-depth behavior.

\begin{figure}[t!]
    \includegraphics[width=\columnwidth]{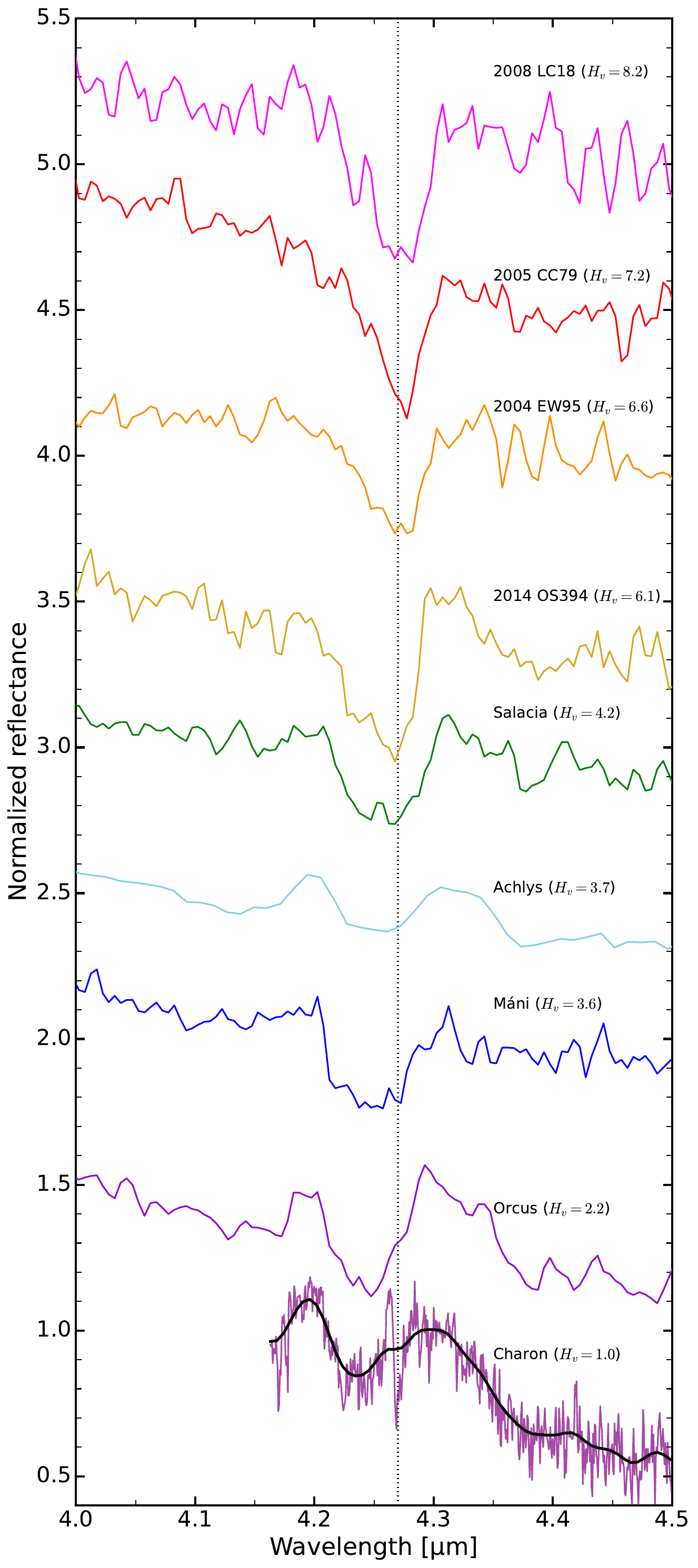}
    \caption{A selection of 4.0--4.5\,\um\ reflectance spectra of \water-type KBOs, normalized to unity at 4.3\,\um. The spectra are arranged from top to bottom in order of increasing object size. The vertical dotted line at 4.27\,\um\ denotes the expected band center for crystalline \cotwo\ ice. The larger objects display distinctly broader and blue-shifted absorption features than the smaller KBOs. The spectrum of Charon, obtained with the high-spectral-resolution G395H grating of NIRSpec, reveals complex fine structure within the \cotwo\ band. The black curve shows the same spectrum convolved to the lower spectral resolution of the PRISM grating.}
    \label{fig:co2}
    \vspace{-0.2cm}
\end{figure}

While it may be tempting to attribute the observed transition in \cotwo\ band depth to a systematic difference in relative \cotwo-ice abundance, a recent compositional analysis of Charon's JWST spectrum has highlighted the complexities inherent in interpreting \cotwo-ice features in spectra dominated by \water-ice \citep{protopapa2024}. Charon's high-resolution spectrum ($\Delta\lambda/\lambda \sim 2700$) reveals the narrow $\nu_1+\nu_3$ combination band at 2.7\,\um\ in addition to the $\nu_3$ fundamental at $\sim$4.25\,\um, enabling more robust characterization of the physical nature of \cotwo. In that study, the position and shape of the $\nu_1+\nu_3$ combination band at 2.7\,\um\ were attributed with high confidence to crystalline \cotwo\ segregated from \water, while the form of the $\sim$4.25\,\um\ fundamental band was attributed to a two-layer structure in which amorphous \water\ ice and crystalline \cotwo\ particles overlie a substrate of crystalline \water\ ice and tholin-like material. An interplay between endogenous and exogenous sources of \cotwo\ was invoked, with the preferred explanation involving subsurface \cotwo\ exposed by cratering events \citep{protopapa2024}. For the other objects in our \water-type KBO sample, the \cotwo\ combination band at 2.7\,\um\ would be greatly attenuated at the low resolution ($\Delta\lambda/\lambda \sim 100$) of the spectra. To roughly quantify the detectability of this feature on \sal\ and \man, we fit the 2.6--2.8\,\um\ continuum with a cubic spline and constrained the relative magnitude of reflectance deviations near 2.7\,\um\ to at most a few percent.

A close inspection of the \cotwo\ feature across the ensemble of \water-type KBO spectra reveals a notable variation in the $\sim$4.25\,\um\ band shape. Figure~\ref{fig:co2} plots a selection of spectra for objects with a range of $H_{V}$ values. While the relatively small objects show a characteristic V-shaped absorption feature with a minimum near 4.27\,\um---consistent with pure crystalline \cotwo\ ice \citep[e.g.,][]{gerakines2020} or \water:\cotwo\ mixtures \citep[e.g.,][]{protopapa2024,depra2025}---the larger KBOs display broader signatures with flatter reflectance minima that extend to shorter wavelengths. The \cotwo\ feature in the spectrum of Charon, obtained at a significantly higher spectral resolution, exhibits a unique double-peaked absorption profile, with two minima separated by a reflectance peak at 4.265\,\um. Comparisons with the JWST spectrum of Europa, which presents a similar although not identical double-lobed absorption band \citep{trumbo2023,villanueva2023}, together with laboratory spectra of \cotwo\ ice preirradiation and postirradiation in both pure and mixed configurations, have enabled investigation into whether Charon’s distinctive \cotwo\ band structure is due to separate contributions from pure crystalline \cotwo\ ice and \cotwo\ embedded in a complex \water/organic-rich molecular environment \citep{protopapa2024}. An alternative explanation, which was favored by the authors and motivated by the layered configuration containing a fine-grained upper layer of \cotwo\ ice, posits that the apparent double-peaked structure is caused by a Fresnel peak of \cotwo\ ice \citep{protopapa2024}.

Looking farther afield to other small body populations, we find numerous examples of blue-shifted \cotwo\ fundamental absorption bands among the Jovian and Saturnian moons, including Callisto, Ganymede, Iapetus, Phoebe, and several irregular satellites \citep{hibbitts2000,hibbitts2003,cruikshank2010,pinillaalonso2011,bockeleemorvan2024,cartwright2024,belyakov2025,sharkey2025}. A similar 4.25\,\um\ \cotwo\ feature is also seen on Eurybates---a collisional family member within the Jupiter Trojans \citep{wong2024}. Due to the higher surface temperatures of these middle solar system bodies, pure \cotwo\ ice is unstable and must therefore be present in more refractory configurations, e.g., \water--\cotwo\ mixtures, mineral-bound \cotwo, or irradiation-produced carbonic acid \citep[e.g.,][]{hibbitts2003,jones2014,schiltz2024}.

The limitations in the resolving power and signal-to-noise ratio in the sample of KBO spectra obtained with the prism prevent us from being able to make definitive statements about the nature of \cotwo\ ice on these bodies. Convolving the spectrum of Charon to the same resolving power as the other spectra largely obscures the fine structure, as illustrated by the solid black curve in Figure~\ref{fig:co2}. Moreover, as mentioned previously, we do not detect \cotwo\ overtone bands around 2.7\,\um, which have been shown to be highly diagnostic of the particular configuration of \cotwo\ on the surface---whether the \cotwo\ is in a pure state \citep[e.g.,][]{raut2013,gerakines2020} or incorporated into a mixed medium with other ices \citep[e.g.,][]{ehrenfreund1999,protopapa2024}. Nevertheless, the broad agreement in overall \cotwo\ fundamental band shape between Charon and the other bright KBOs---including \sal\ and \man---may reflect similar underlying spectral profiles, suggesting that the same processes that drive the complex \cotwo\ spectral signature on Charon may be at play across the large-sized and midsized KBO population. Such a scenario could be confirmed through the analysis of higher-quality spectra derived from future observations, combined with additional laboratory investigations into the effects of thermal and radiation-driven processing on \cotwo\ ice.


\section{Summary}
\label{sec:concl}

We presented the JWST/NIRSpec spectra of the midsized KBOs \sal\ and \man. The two spectra show clear similarities, with prominent absorption bands of \water\ and \cotwo\ ice. No additional features from CH$_{4}$, C$_{2}$H$_{6}$, or other hydrocarbons were detected. A comparison with spectroscopic data of smaller objects in the trans-Neptunian region indicates that \sal\ and \man\ belong to the spectrally prominent water (\water-type) KBO class. An ensemble analysis of \water-type spectra revealed several salient trends within this population:\vspace{-0.1cm}
\begin{enumerate}
    \item The band depth of the \water-ice absorption feature at 3\,\um\ correlates strongly with object size, as inferred from the absolute magnitudes, with larger KBOs displaying deeper bands. This covariance suggests an overall increase in the relative surface abundance of \water\ ice, although additional variability across the sample due to grain size differences may also contribute to the distribution of band depths.\vspace{-0.1cm}
    \item While an analogous positive covariance between the measured 2\,\um\ \water-ice band depths and target diameter is evident among the largest KBOs, this trend becomes less pronounced at smaller sizes. There is strong evidence for a bimodality in 2\,\um\ band depth among KBOs with $4 < H_{V} < 8$, suggesting systematic differences in \water-ice grain size or vertical stratification among objects with similar sizes.\vspace{-0.1cm}
    \item There is a marked transition in the strength of the $\sim$4.25\,\um\ \cotwo-ice absorption band at $H_V \sim 5-6$. Larger objects have systematically shallower \cotwo\ features, although the presence of selection biases within the spectroscopic sample with respect to the various KBO dynamical classes prevent us from making a definitive claim about the size-dependent nature of this apparent trend.\vspace{-0.1cm}
    \item The shape of the $\sim$4.25\,\um\ \cotwo-ice feature also appears to vary with target size. While smaller objects have bands centered at $\sim$4.27\,\um\ that are consistent with pure crystalline \cotwo\ ice or \water:\cotwo\ mixtures, the larger KBOs within our sample show broader features that extend to shorter wavelengths. The latter may be caused by additional contributions from either (1) irradiated and/or complexed \cotwo, or (2) a thin veneer of \cotwo\ overlaying a reflective \water-rich layer, as suggested by previous spectral modeling of Charon.\vspace{-0.1cm}
\end{enumerate}

The trend between \water-ice abundance and size provides tentative evidence for the emergent effects of internal heating and cryovolcanic activity on midsized KBOs such as \sal\ and \man, although follow-up investigations are needed to disentangle the effects of \water-ice grain size and possible vertical stratification on the surfaces. Future work that synthesizes these results with new insights from studies of interior thermal evolution, investigations of endogenous and exogenous surface alteration, and detailed spectral modeling will greatly refine our understanding of the various physical and chemical processes that have sculpted the observed compositional diversity among water-ice-rich KBOs.


\section*{Acknowledgments}
This work is based on observations made with the NASA/ESA/CSA James Webb Space Telescope. The data were obtained from the Mikulski Archive for Space Telescopes at the Space Telescope Science Institute, which is operated by the Association of Universities for Research in Astronomy, Inc., under NASA contract NAS 5-03127 for JWST. These observations are associated with Program \#1191. The specific observations analyzed can be accessed via \dataset[https://doi.org/10.17909/252c-8q63]{https://doi.org/10.17909/252c-8q63}. 

H.B.H. and S.N.M. acknowledge support from NASA JWST Interdisciplinary Scientist grant 21-SMDSS21-0013. R.B. acknowledges support from the CNES-France (JWST mission). N.P.-A. acknowledges funding through the ATRAE program of the Ministry of Science, Innovation, and Universities (MCIU) and the State Agency for Research (AEI) in Spain.

\facilities{JWST/NIRSpec.}
\software{\texttt{astropy} \citep{astropy2013,astropy2018,astropy2022}, \texttt{jwst} \citep{jwst}, \texttt{jwstspec} \citep{jwstspec}, \texttt{matplotlib} \citep{matplotlib}, \texttt{numpy} \citep{numpy}, \texttt{scipy} \citep{scipy}.}


{\bibliography{main.bib}{}
\bibliographystyle{aasjournalv7}}

\end{document}